\documentclass[manuscript,nonacm,margin=0.5in]{acmart}
\pdfoutput=1
\usepackage[utf8]{inputenc}
\usepackage{multicol}
\usepackage{geometry}
\usepackage[backend=biber,style=numeric,sorting=none]{biblatex}
\usepackage{graphicx}
\graphicspath{ {./images/} }
\usepackage{multirow}
\usepackage{float}

\begin{document}

\title[Addressing Bias in Visualization Recommenders by Identifying Trends in Training Data]{Addressing Training Bias in Machine Learning Visualization Recommendation Systems by Identifying Trends in Training Data}
\subtitle{Improving VizML Through a Statistical Analysis of the Plotly Community Feed}

\author{Allen Tu}
\email{atu1@umd.edu}
\affiliation{%
  \institution{University of Maryland, College Park}
  \city{College Park}
  \state{MD}
  \country{USA}
}
\author{Priyanka Mehta}
\email{pmehta1@umd.edu}
\affiliation{
  \institution{University of Maryland, College Park}
  \city{College Park}
  \state{MD}
  \country{USA}
}
\author{Alexander Wu}
\email{wu3567@umd.edu}
\affiliation{
  \institution{University of Maryland, College Park}
  \city{College Park}
  \state{MD}
  \country{USA}
}
\author{Nandhini Krishnan}
\email{nankri3@umd.edu}
\affiliation{%
  \institution{University of Maryland, College Park}
  \city{College Park}
  \state{MD}
  \country{USA}
}
\author{Amar Mujumdar}
\email{amarmuju@umd.edu}
\affiliation{%
  \institution{University of Maryland, College Park}
  \city{College Park}
  \state{MD}
  \country{USA}
}

\renewcommand{\shortauthors}{Tu, Mehta, Wu, Krishnan, and Mujumdar}

\maketitle

\begin{multicols}{2}
\section*{Abstract}
Machine learning is a promising approach to visualization recommendation due to its high scalability and representational power. Researchers can create a neural network to predict visualizations from input data by training it over a corpus of datasets and visualization examples. However, these machine learning models can reflect trends in their training data that may negatively affect their performance. Our research project aims to address training bias in machine learning visualization recommendation systems by identifying trends in the training data through statistical analysis.

\section{Introduction}

Our project focuses on VizML \cite{10.1145/3290605.3300358}, a machine learning recommendation system that is trained over 2.3 million dataset-visualization pairs from the open-source Plotly Community Feed. VizML is adept at providing interpretable measures of feature importance, and its models can easily be integrated into existing visualization systems due to their learned understanding of design choices. VizML recommendations are quantifiably similar to visualizations created by human analysts, and its performance exceeds that of other visualization recommendation systems.

However, VizML’s creators acknowledge that it suffers from significant training bias. In other words, VizML models reflect trends in the Plotly Community Feed that heavily influence its performance. The authors do not elaborate any further, and we did not find any existing discussions about trends in the Plotly Community Feed. Furthermore, bias in machine learning recommendation systems seems to be an underexplored topic as a whole. 

The sheer size of VizML models poses several challenges for offsetting their bias. Since we are limited in both time and computing power, we are not able to experiment by training multiple full VizML models. We cannot create a similar dataset for testing, and it is difficult to adapt other open-source datasets to the required input format. Additionally, the creators of VizML do not explain their findings about training bias, so we do not have any prior knowledge of the trends that we are searching for. Thus, performing a statistical analysis of the Plotly Community Feed is our optimal approach.

In this paper, we perform a statistical analysis to identify trends in the training data. Then, we evaluate the significance of these trends and connect them to the performance of the model. Finally, we develop methods to address our findings and improve the model. Although our results are specific to VizML, our process is generalizable and can be used to improve any machine learning visualization recommendation model.

Our insight is that we can address training bias in machine learning visualization recommendation systems by identifying trends in the training data through statistical analysis. In our findings, we identified several trends in the Plotly Community Feed that influence recommendations by the VizML model. Some of these trends are conscious design decisions that improve the readability of visualizations, while others demonstrate how Plotly users tend to create charts based on convenience rather than effectiveness. We can use our findings to improve the model by counteracting detrimental trends and improving in areas of deficiency. 

The next section of our paper will focus largely on related work. We discuss machine learning visualization recommendation systems, the VizML model, and the current state of research on its training bias. Then, we detail our overall approach and how we collected and processed our data. Our findings are centered around our statistical analysis of the Plotly Community Feed. We identify significant trends in the data and connect them to the performance of the model. In our conclusion, we summarize our results and explore avenues for addressing the training bias.

\section{Related Work}

Visualizing data is often the first step of analysis; however, visualization tools can be difficult and tedious to use because they require the user to manually specify options via code or menus. Recent computer science research has led to sophisticated systems that automatically suggest tailored visualizations based on the selected data and the user’s needs. These recommenders can make creating visualizations more accessible to people without an analytical background as well as assist analysts in producing them more efficiently.

Current visualization recommenders, such as VisC \cite{10.1007/978-3-319-07233-3_58} and Voyager 2 \cite{2016-voyager}, rely on large sets of predefined questions and rules. While these systems are effective to an extent, they are costly to create in terms of labor because they require domain experts to manually write constraints. Additionally, this process hinders the system’s ability to identify trends, incorporate contextual information, and adapt to user behavior \cite{10.1145/3092931.3092937}. 

A machine learning approach solves many of these limitations. Instead of relying on a set of manually specified constraints, a machine learning model generates predictions based on training data, so improvements can be made relatively easily by retraining the model or fine-tuning its design. These neural networks are also much more effective at capturing complex relationships and can be trained on vast amounts of high-dimensional data. 

\subsection{VizML}
{
  \centering
  \includegraphics[width=\linewidth]{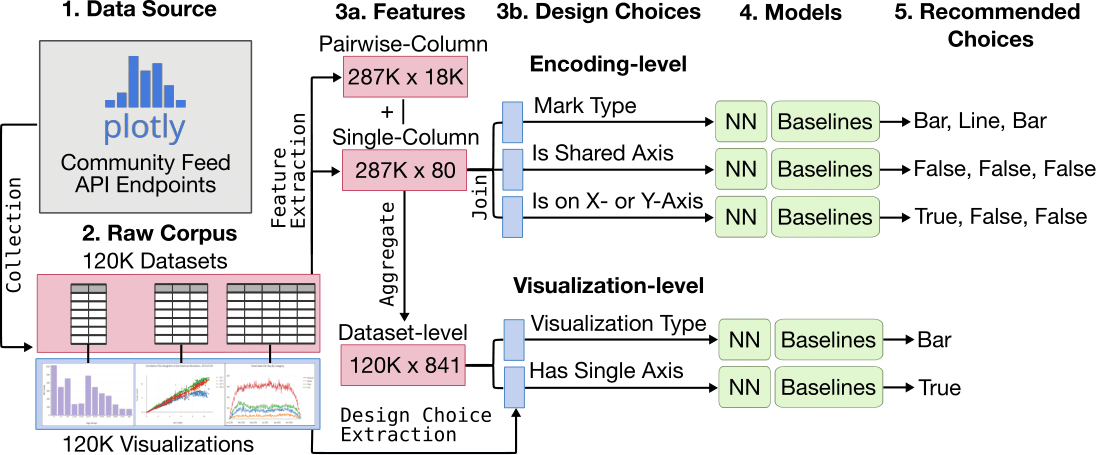}
  \captionof{figure}{“Diagram of data processing and analysis flow in VizML, starting from (1) the original Plotly Community Feed API endpoints, proceeding to (2) the deduplicated dataset-visualization pairs, (3a) features describing each individual column, pair of columns, and dataset, (3b) design choices extracted from visualizations, (4) task-specific models trained on these features, and (5) potential recommended design choices” \cite{10.1145/3290605.3300358}.}
}

VizML \cite{10.1145/3290605.3300358}, a more recent project, learns the relationships between datasets and the visualizations analysts create from them by training on 2.3 million dataset-visualization pairs from the Plotly Community Feed. VizML models are adept at providing interpretable measures of feature importance and can easily be integrated into existing visualization systems. They are also trained on an extremely large and diverse corpus that was created based on real visual analysis by analysts on their own datasets, so their representational ability far exceeds that of older recommenders like Draco-Learn \cite{Draco} and DeepEye \cite{DeepEye}. 

However, it is not without limitations. VizML’s creators acknowledge that the model suffers from significant training bias. In other words, VizML models reflect trends in the Plotly Community Feed that may significantly affect its performance. The authors do not elaborate any further, and we did not find any existing discussions about trends in the Plotly Community Feed. Furthermore, bias in machine learning recommendation systems seems to be an underexplored topic as a whole. Our research project aims to provide insights into VizML’s training bias through a statistical analysis of the training data. 

\section{Approach}
The goal of our project is to address training bias in the VizML model. We begin by collecting and processing training data from the Plotly Community Feed. Then, we conduct a statistical analysis of the dataset-visualization pairs by evaluating the frequency of each chart type. We discover and interpret trends in the training data and connect them to the training bias of the VizML model. Finally, we explore methods for addressing VizML’s training bias based on our findings, avenues for further statistical analysis, and possible expansions for the VizML model in our conclusion. These steps are generalizable and can be used to improve any machine learning visualization recommendation system.

Our insight is that we can address training bias in machine learning visualization recommendation systems by identifying trends in the training data. In our statistical analysis, we identified several trends in the Plotly Community Feed that influence recommendations by the VizML model. Some of these trends are conscious design decisions that improve the readability of visualizations, while others demonstrate how Plotly users tend to create charts based on convenience rather than effectiveness. We can use our findings to improve the model by counteracting detrimental trends and improving in areas of deficiency. 

\subsection{Data Collection}

The VizML model is trained on the Plotly Community Feed dataset, which consists of over 2.3 million dataset-visualization pairs. Analyzing the entire corpus was unrealistic given our available time and computing power, so we acquired a 1,000 example subset of the training data from the VizML GitHub repository. Then, we loaded the data into a Pandas DataFrame inside of a Jupyter Notebook for exploration and statistical analysis.

{
  \centering
  \includegraphics[width=\linewidth]{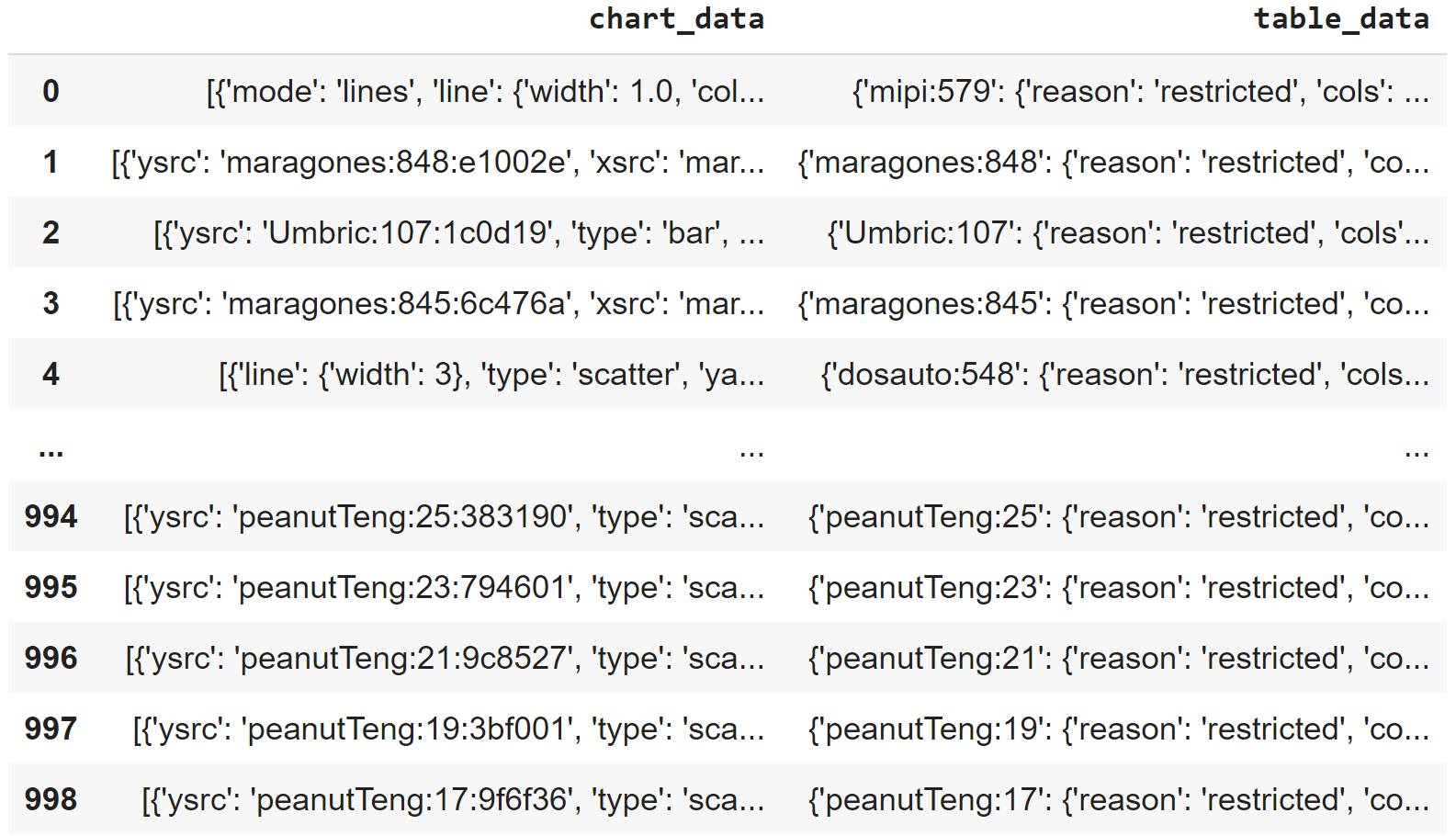}
  \captionof{figure}{Pandas DataFrame of the processed dataset.}
}

We found that the dataset is organized into four columns. \textit{fid}, the example’s unique identifier, and \textit{layout}, the formatting specifier of the chart, are not relevant to our analysis, so we discarded them from the DataFrame entirely. The remaining two columns are \textit{chart\_data}, which specifies the user-created chart, and \textit{table\_data}, which contains the user-submitted data that the chart was created from. We convert each entry in these columns from the JSON format into lists and dictionaries that are easily interpreted by most Python packages and libraries.

{
  \centering
  \includegraphics[width=2.41in]{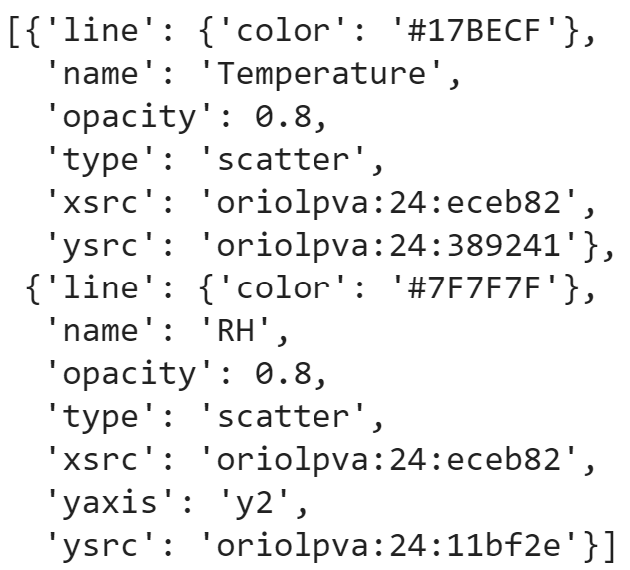}
  \captionof{figure}{Scatter plot in \textit{chart\_data}. ‘Temperature’ and ‘RH’ are plotted across a single predictor variable. Regression lines for each response variable are drawn over the data points.}
}

Each row in \textit{chart\_data} is a list of chart objects represented by dictionaries. The entries in the dictionary specify the properties of the chart object. All dictionaries are required to include the \textit{name} of the object, the \textit{type} of chart, and the \textit{xsrc} and \textit{ysrc} of the predictor and response variables, respectively, in the corresponding \textit{table\_data} entry. The user can also specify optional fields. For example, they can fix the \textit{color} of the object to turquoise, or they can update \textit{mode} to add an automatically calculated regression line to a ‘scatter’ \textit{type} object.

\subsection{Statistical Analysis}

The vast majority of the \textit{chart\_type} objects across our entire 1,000 example dataset only have the four required fields: \textit{name}, \textit{type}, \textit{xsrc}, and \textit{ysrc}. \textit{name} is not useful for our purposes because VizML generates them based on the column labels in the data. VizML also relies on the user to select the data query, so it is difficult to measure the impact of \textit{xsrc} and \textit{ysrc} on model bias through only a statistical analysis of the training data. The remaining field is \textit{type}, which most significantly influences the VizML model because it specifies the type of chart that is created from the input data query.

By analyzing the trends in the \textit{type} field, we will be able to draw insights regarding the types of visualizations that the VizML model is likely to recommend. Our statistical analysis focuses on the overall frequency and complexity of charts typical of our subset of the Plotly Community Feed. We determined the frequency of each of the 15 chart types in our dataset and identified trends in the preferences of Plotly users. For each of the three most common chart types, we evaluated trends in their complexity by measuring the number of response variables or categories visualized by each of their corresponding examples.
\section{Findings}

\subsection{Frequency of Chart Types}

First, we calculate the frequency of each type of chart in our dataset. Each row in \textit{chart\_data} maps to exactly one chart type. It is worth noting that line charts are initialized with the \textit{line} field instead of the \textit{type} field, and they are distinct from scatter plots with a line overlaid on the point distribution. Charts initialized without a \textit{type} field default to scatter plots.

{
  \centering
  \includegraphics[width=\linewidth]{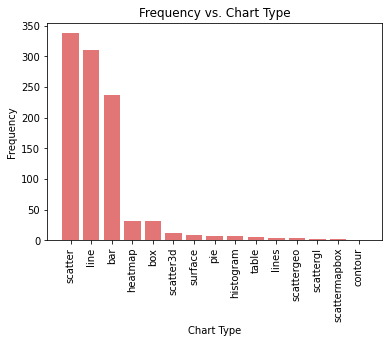}
  \captionof{figure}{Frequency of each chart type in the dataset. Scatter, line, and bar plots account for 39.6\%, 31.0\%, and 17.9\% of the examples, respectively. The remaining twelve less common chart types account for 11.4\% of the examples.}
}

We find that scatter plots, line plots, and bar plots are by far the most common chart types in the Plotly Community Feed. We expected this because these chart types are simple and effective at visualizing trends in data, and they are often the first visualizations that analysts reach for in their initial exploration of the data. VizML is almost guaranteed to recommend one of these three chart types for every selected data query. 

It is unlikely for the model to ever recommend any of the other twelve chart types, which  account for only 11.4\% of the dataset all together. For example, the fourth most common chart type is the heatmap, which has a relative frequency of 0.031. The scatter plot has a relative frequency of 0.396, so it is almost 13 times more likely for a Plotly user to create a scatter plot than a heatmap. Since many of these chart types fill somewhat niche roles, this is also expected. 

However, a notably underrepresented chart type is the histogram. It is a ubiquitous, effective chart type that is commonly used by analysts to visualize datasets typical of the Plotly Community Feed. Most analysts would likely expect its relative frequency to be much higher than 0.006; in fact, we expected it to be the fourth most common chart type.

A possible explanation is that a histogram can be relatively difficult to set up in Plotly. If the user is not satisfied with the automatically generated bins, they have to specify them manually. This process requires some deeper knowledge of the data, and additional coding is needed to manually specify bins over a wide range. Thus, Plotly users may find it more convenient to choose a scatter or line plot instead, even if they believe a histogram may be more effective at visualizing their data.

\subsection{Analysis of Scatter Plots}

{
  \centering
  \includegraphics[width=\linewidth]{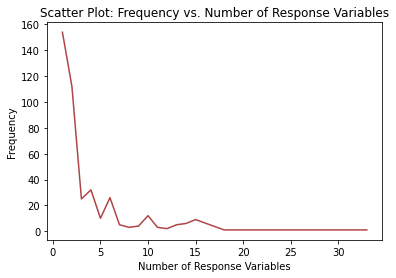}
  \captionof{figure}{Frequency of the number of response variables across all scatter plots in the dataset. One and two response variables account for 37.1\% and 27.0\% of the examples, respectively. We discarded the extreme outliers 61 and 216 response variables for readability.
}

}
{
  \centering
  \includegraphics[width=\linewidth]{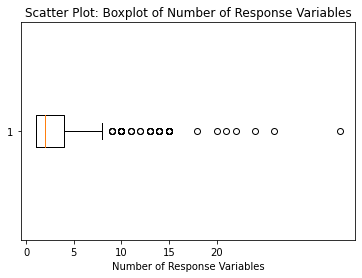}
  \captionof{figure}{Boxplot of the number of response variables across all scatter plots in the dataset. The quartiles of the distribution are 1, 2, and 4 response variables. 
}
}

Next, we examined the number of objects in each chart to gauge their complexity. Each object in a scatter plot represents a different response variable. We found that 64.1\% scatter plots in the Plotly Community Feed have one or two response variables, but it is normal for them to have up to four. It is unlikely for scatter plots to have five or more response variables, although an extreme outlier that was omitted from the visualizations has 216. 

The preference clearly leans towards simple scatter plots with less response variables. Plotly users are unlikely to select more than four response variables even if more data is available. A possible explanation is that a scatter plot can become unreadable if the density of datapoints is too high, so using too many response variables damages the effectiveness of the visualization. Additionally, the \textit{plotly.scatter} method creates single response variable scatter plots by default, so users may prefer creating these simple scatter plots out of convenience.

\subsection{Analysis of Line Plots}
{
  \centering
  \includegraphics[width=\linewidth]{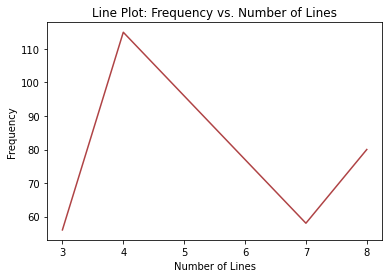}
  \captionof{figure}{Frequency of the number of lines across all line plots in the dataset.  3, 4, 7, and 8 lines account for 18.1\%, 37.2\%, 18.8\%, and 25.9\% of the examples, respectively. We discarded the extreme outlier 50 lines for readability.
}
}
{
  \centering
  \includegraphics[width=\linewidth]{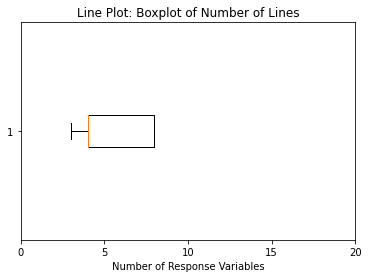}
  \captionof{figure}{Boxplot of the number of lines across all line plots in the dataset. The quartiles of the distribution are 4, 4, and 8 lines. 
}
}

Each object in a line plot represents a single line. We found that most line plots in the Plotly Community Feed have four to eight lines, but an extreme outlier that was omitted from the visualization has 50. Our analysis indicates it is unlikely for Plotly users to create a line plot with only one or two lines; in fact, none were present in our 1,000 example subset. 

In a previous segment of our analysis, we found that many scatter plots with only one or two response variables use the \textit{mode} field to overlay a regression line. A possible explanation is that Plotly users prefer simpler, less cluttered graphs. If a scatter plot has too many response variables, the visual density of the data points and regression lines may make it difficult to read. To solve this problem, the user can omit the data points entirely by creating a line plot instead. Our analysis provides statistical evidence that supports the theory that Plotly users prefer scatter plots for visualizing one or two lines and line plots for visualizing three or more lines. 

\subsection{Analysis of Bar Plots}

{
  \centering
  \includegraphics[width=\linewidth]{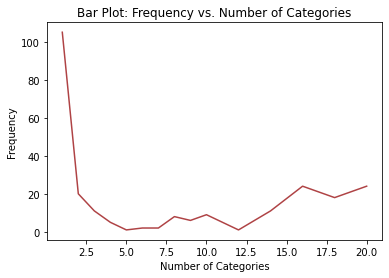}
  \captionof{figure}{Frequency of the number of categories across all bar plots in the dataset. 1 category accounts for 42.5\% of the examples, while 2, 16, 18, and 24 categories account for 8.1\%, 9.7\%, 7.3\%, and 9.7\% of the examples, respectively.
}
}
{
  \centering
  \includegraphics[width=\linewidth]{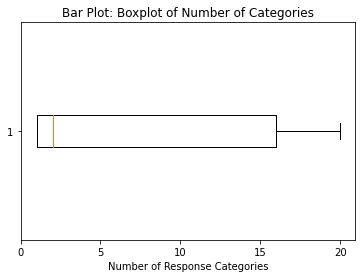}
  \captionof{figure}{Boxplot of the number of categories across all barplots in the dataset. The quartiles of the distribution are 1, 2, and 16 categories.
}
}

Each object in a bar plot represents a single category, or bar. 105, or 42.5\%, of the bar plots only have one category. We found this to be abnormal because a primary purpose of a bar plot is to visually compare different categories. The representational capacity of a single bar is limited because there is no other data to compare it to. One category bar graphs essentially represent only a single number or proportion, and most analysts would agree that it is more effective to use another format to represent it. Thus, the high relative frequency of single category bar graphs is a significant indicator of potential bias in the Plotly Community Feed.

Interestingly, our distribution also indicates that it is just as common for Plotly users to visualize two categories as it is for them to visualize 16 to 20 categories. Upon further statistical analysis, we determined that Plotly users tend to create bar plots that either visualize a single category or every available category in their table. A possible explanation is that these are the two most convenient approaches when using the \textit{plotly.bar} method. The user can select one \textit{xsrc} and \textit{ysrc} pair to create a single bar, or they can select their entire dataset and quickly create a bar for every category.

A more powerful visualization often lies in the middle ground. For example, it may be more effective to show only the three most significant categories with over 250 counts each without the other 17 categories with less than 10 counts each. We found that bar plots with two to seven categories are more likely to have selected only a subset of the available categories in their table. This requires some preliminary knowledge of the dataset as well as additional coding, which may explain why these examples are relatively less frequent. Therefore, our statistical analysis suggests that users who picked one or all of the categories may have done so out of convenience rather than effectiveness. 

\section{Conclusion}

We found that the Plotly Community Feed heavily prefers a small subset of the available chart types. Data queries with one to two response variables are nearly always used to create scatter plots, while those with three to seven response variables are sometimes used in line plots. Categorical data queries are almost exclusively used to create bar plots. Since it uses the Plotly Community Feed as training data, VizML strongly reflects these trends. Recommendations made by the model given any data query are almost guaranteed to be scatter, line, or bar plots; it is unlikely for the remaining twelve chart types to be recommended at all.

VizML’s significant bias toward charts typical of its training data is not necessarily negative. It recommends scatter, line, and bar plots because they are the essential charts that the vast majority of Plotly users would only create. Some preferences, such as choosing between line and scatter plots based on the number of response variables, are practical design choices that can increase the effectiveness of generated charts. Recommendations by VizML are quantifiably similar to visualizations created by human analysts, and the model’s performance exceeds that of other visualization recommendation systems. 

{
  \centering
  \includegraphics[width=3in]{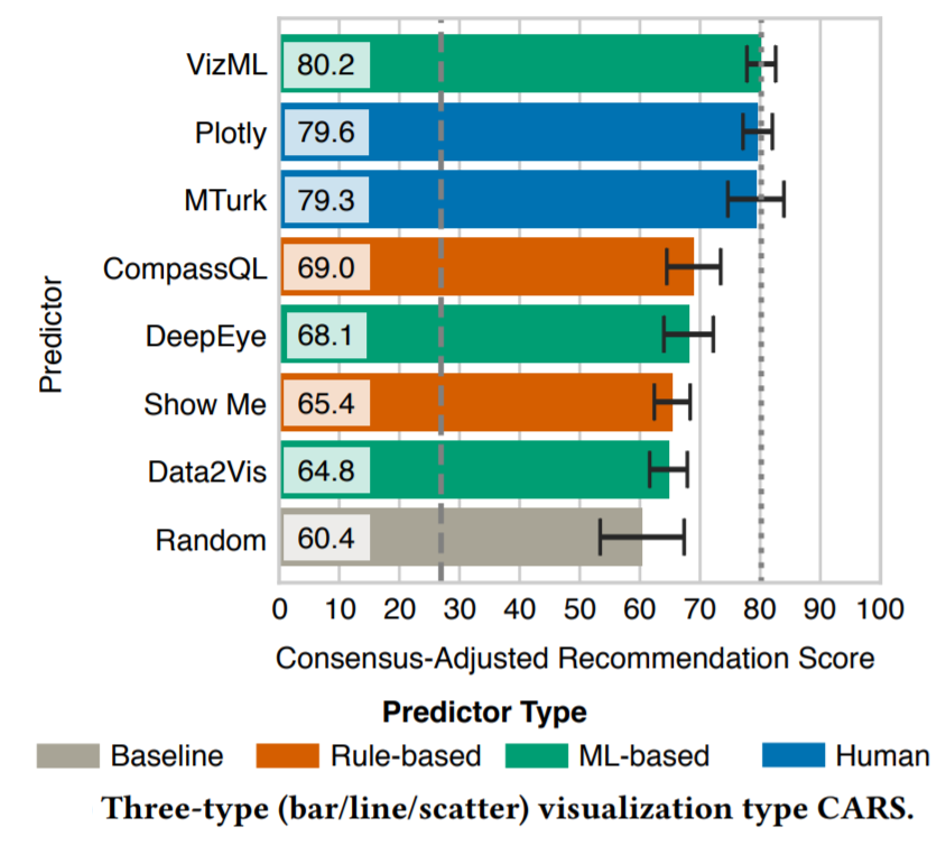}
  \captionof{figure}{“Consensus-Adjusted Recommendation Score of three ML-based, two rule-based, and two human predictors when predicting consensus visualization type. Error bars show 95\% bootstrapped confidence intervals, with 105 bootstraps. The mean minimum achievable score is the lower dashed line, while the highest achieved CARS is the upper dotted line” \cite{10.1145/3290605.3300358}.
}
}

However, accurately predicting human behavior is not always positive either. For example, the histogram is a common and effective chart type, but it is severely underrepresented in our dataset. Since histograms are relatively difficult to set up in Plotly, users may choose to use a scatter or line plot instead, even if they believe a histogram would be more effective. Similarly, we found that 42.5\% of bar plots in the dataset visualize a single category, limiting their representational power. Further analysis also revealed that users tend to create bar plots with either one or every category – the two approaches that are the fastest to perform in Plotly.

Our insight is that some trends in the Plotly Community Feed are the result of convenience rather than effectiveness. Scatter, line, and bar plots are heavily preferred because they are common charts that are easy and simple to create. Histograms can require more setup than line plots, so they are neglected almost completely. Users tend to choose one or all of the categories when creating a bar plot because determining which categories are significant requires an intermediary step of analysis. 

The goal of a visualization recommendation system is to recommend a high quality visualization from a given data query, but the VizML model is influenced by how easy or convenient it is to make certain visualizations in Plotly. The user does not save any time because the recommendation process is automatic, and the generated visualization may have been of higher quality if there were more training examples that required more effort to manually specify. Therefore, VizML’s training bias can be detrimental because human behavior is influenced by factors beyond optimizing performance. Addressing these weaknesses in the VizML model may lower its score on consensus-adjusted evaluators; however, the resulting visualizations recommended by the model will ultimately be of higher quality. 

\subsection{Avenues for Further Analysis}

10,000 and 100,000 example subsets of the Plotly Community Feed, as well as the entire 2.3 million example dataset, are available on the VizML GitHub repository. Evaluating these larger datasets could reveal trends beyond the ones we discovered in our 1,000 example subset. The scope of our statistical analysis can also be extended to include \textit{layout} and \textit{table\_data}, as well as the remaining fields in \textit{chart\_data}. Some of the relationships in these areas only affect a small demographic of charts, but they may have implications that relate to our findings. 

Due to our limited time and computing resources, we were not able to perform experiments using the VizML model itself. A promising expansion is to exaggerate different characteristics in the training data. Training VizML using these variations could provide insight into how to affect biases in the model. For example, experimenting with the proportion of chart types may address the underrepresentation of histograms. Another experiment could measure the performance of VizML on datasets that are not typical of the Plotly Community Feed to identify areas for improvement.

\subsection{Improving VizML}
Our statistical analysis concluded that the Plotly Community Feed contains a large number of examples that may have been influenced by convenience rather than effectiveness. For example, the proportion of bar plots that visualize a single category is abnormally high. We can identify and clean examples with potentially unwanted characteristics like these, reducing their frequency in the training data and their influence on the VizML model. They can easily be replaced with better unused examples from the Plotly Community Feed.

Some chart types such as histograms are underrepresented by the training data and will not be recommended by VizML. We can increase their relative frequency by adding unused examples with these chart types from the Plotly Community Feed. Since the training data is no longer a random sample of the population, experimentation is necessary for finding the appropriate number of additional examples. The new model will recommend these previously underrepresented chart types, so the training bias is in part addressed. If VizML is configured to recommend multiple visualizations, this will also help diversify the chart types in the output.

Many chart types, like heatmaps and contour plots, fill niche purposes and are ill-suited for a general use recommender like VizML. Fortunately, their combined effect on the model’s performance is negligible due their low relative frequency. A possible avenue for expansion is to train a VizML model to specifically recommend these more complex visualizations. Approximately 7\% of our 1,000 example subset use these chart types, so a sizable training dataset could be built by selecting all useful examples from a larger subset.

\printbibliography
\end{multicols}
\end{document}